\documentclass[journal,article,accept,moreauthors,12pt,a4paper]{mdpi} 
\lastpage{1783}
\doinum{10.3390/e14091771}
\pubvolume{14}
\pubyear{2012}
\history{Received: 21 August 2012; in revised form: 13 September 2012 / Accepted: 17 September 2012 / \\
Published: 20 September 2012}

\usepackage{graphicx}
\usepackage{amsmath,amssymb}
\usepackage{graphicx,color}
\usepackage{multirow}
\usepackage{soul}
\def\bea{\begin{eqnarray}}
\def\eea{\end{eqnarray}}

\def\beq{\begin{equation}}
\def\eeq{\end{equation}}

\newbox\pippobox

\input epsf
\Title{Exact Solution and Exotic Fluid in Cosmology}

\Author{Seyen Kouwn $^{1}$,
Taeyoon Moon $^{2}$
and Phillial Oh $^{1,}$*}

\address{%
$^{1}$ Department of Physics and Institute of Basic Science,
 Sungkyunkwan University, Suwon 440-746, Korea;
E-Mail: seyen@skku.edu \\

$^{2}$ Center for Quantum
Space-time, Sogang University, Seoul 121-742, Korea; \linebreak
E-Mail:~dpproject@skku.edu}

\corres{E-Mail: ploh@skku.edu; \linebreak{Tel./Fax: 82-031-290-7046/82-031-290-7055}}

\abstract{We investigate  cosmological consequences of nonlinear sigma model
coupled with a cosmological fluid which satisfies the continuity
equation. The target space action is of the de Sitter type and is
composed of four scalar fields. The potential which is a function of
only one of the scalar fields is also introduced. We perform a
general analysis of the ensuing cosmological equations and give
various  critical points and their properties. Then, we show that
the model exhibits an exact cosmological solution which yields a
transition from matter domination into dark energy epoch and compare it
with the  $\Lambda$-CDM behavior. Especially, we calculate the age of
the Universe and  show that it is consistent with the observational
value if the equation of the state $\omega_f$ of the cosmological
fluid is within the range of $0.13 < \omega_f < 0.22.$ Some
implication of this result is also discussed.}


\begin{document}


\section{Introduction}

The recent astronomical measurements
revealed that
the current Universe is accelerating \cite{Riess:1998cb,Riess:1998cb2}.
It is believed that the acceleration is caused
by an unknown energy, \emph{i.e}., dark energy, and grasping the identity of dark energy
 is one of most fundamental problems in the modern cosmology
\cite{mukh}. Many theoretical models for dark energy
have been proposed ever since.
Among them, the standard approach is to
introduce a cosmological constant \cite{wein,wein2,wein3,wein4,wein5,wein6}.
In spite of its simplicity and theoretical diversities,
it confronts
the extreme fine tuning
problem.
An attractive alternative is to consider that the acceleration is
driven by a scalar field. The well known models include quintessence
\cite{quintessence,quintessence2}
phantom \cite{phantom},
\mbox{k-essence \cite{k-essence,k-essence2,k-essence3}} with the non-canonical kinetic term for the scalar field,
and  quintom \cite{quintom} models.
In these models various aspects of dark energy can be well described in terms of dynamics of the scalar field
and especially, the smallness of the
cosmological constant is attributed to the decaying
scalar energy density \cite{nojiri1}.

On the other hand, dark matter plays a central role in the early
Universe in the process of structure formation \cite{mukh}. Most of
the dark matter is the cold dark matter (CDM) which is
non-relativistic and \mbox{non-baryonic}. In particular, the CDM model with
the cosmological constant $\Lambda$ is established as a standard
cosmological model, $\Lambda$-CDM
\cite{mukh}. This model is in  good agreements with the CMBR \mbox{data
\cite{cmb,cmb2,cmb3}} as well as SN Ia data \cite{Riess:1998cb,Riess:1998cb2}. The special feature of
$\Lambda$-CDM model is that
it assumes the Universe with a zero spatial curvature.

Recently, the dark energy model based on a nonlinear
sigma model \cite{Lee} with or without a cosmological constant was investigated. In this model
the target space is noncompact four-dimensional de Sitter
manifold and four scalar fields are introduced to account for this.
It was found that an exponentially accelerating solution is possible even without the cosmological constant and
that the model could describe dark energy interacting with stiff matter even without any matter present.
The possibility of other kind of dark matter should be also addressed and it motivates to extend to the more general situation
where matter  is introduced separately. Therefore, in this paper, we consider more general case of the de Sitter nonlinear sigma model
where the matter with the equation of
 state $\omega_f$   and also an exponential potential term for the nonlinear sigma model are added to the action, and we investigate the cosmological consequences.

Recall that in spite of the dynamical explanation of the smallness of the current dark energy density contrary to the cosmological constant,
the approaches of ~\cite{quintessence,quintessence2,phantom,k-essence,k-essence2,k-essence3} have their own shortcomings.
For quintessence model,
 it has the nice properties of tracking solution and scaling behavior, but it is somewhat difficult to match the equation of state to be close to $-1$, and has its own fine-tuning problem~\cite{bludman}. The phantom model gives favorable result with the equation of state,
but it has the issue of  quantum instability, even though it is classically stable. The k-essence model has the ghost problem coming from the higher derivative nature of the theory.
We do not attempt to address these difficult issues with the introduction of the four scalar fields, but one advantage is the existence of the exact solution, which makes comparisons with $\Lambda$-CDM more direct. ( See Equations~(\ref{s2}) and (\ref{asol5}).)

Let us mention other salient features of our investigation.
The field content consists of a phantom and the triplet
fields which are canonical scalar fields. Unlike the ordinary
phantom model where Big Rip singularity is known to exist, we have
an explicit solution which does not show such a behavior.
As far as the exact cosmic solution is concerned, the potential
  is of  a negative exponential form  of the
scalar field $\phi$. This solution prolongs from
the matter-dominated epoch to the dark energy epoch.
Recall that in the pure phantom model, the negative
kinetic energy prohibits the potential to assume a negative value.
In our approach, the triplet of scalar fields provides enough compensating positive energy density
such that the weak energy condition is not violated.
This exact solution can be exploited to extract some numerical values that can be compared with current observations.
In particular, we calculate the age of the Universe and compare with the
observations. We find that with a fine-tuning of a couple of
parameters,   $0.13<\omega_f<0.22$ is consistent with the observation.
This matter corresponds to an exotic cosmological fluid.

The paper is organized as follows. In Section II, we present a basic
analysis of $\Lambda-$CDM model paying attention to the exact
cosmological solution and its stability analysis. In  Section III, we
consider the  Einstein gravity coupled with the de Sitter nonlinear
sigma model including an exponential potential and cosmological
fluid with the equation of state $\omega_f$, and discuss the cosmological
evolution equations. In Section IV, the stability analysis is performed
and various critical points are identified. In Section V, we present
the exact cosmological solution for a negative exponential potential
and calculate the current age of the Universe. We compare it with
the observational data and $\Lambda-$CDM model.
Section VI includes conclusion and discussion.

\section{$\Lambda$CDM Model}

Let us first consider an action in which the Einstein gravity has
 a cosmological constant (c.c.) term with a matter term:
\begin{eqnarray}
S=\int \,d^4x \sqrt{-g}~\Big[\frac{1}{2\kappa^2}(R-\Lambda) + {\cal
L}_{matter}\Big]
\end{eqnarray}
where $\kappa^2=8\pi G$. Introducing the standard space-time metric via
\begin{eqnarray}\label{mmetric}
ds^{2}=-dt^2+a^2(t) dx_{i}dx^{i}
\end{eqnarray}
leads to the following equations:
\begin{eqnarray}
H^2&=&\frac{\kappa^2\rho_m}{3}+\frac{\Lambda}{3}\label{h1}\\
\dot{H}&=&-\frac{\kappa^2}{2}(1+\omega_m)\rho_m\label{h15}\\
\dot{\rho}_m&+&3(1+\omega_m)H\rho_m=0\label{h2}
\end{eqnarray}
In these expressions, $\omega_m$ is the equation of state parameter
for the matter term, which satisfies \mbox{$p_{m}=\omega_m\rho_{m}$} when
$T^{\mu}_{\nu}=(-\rho_m,p_{m},p_{m},p_{m})$ are assumed.

In order to check the stability,  we introduce the following dimensionless quantities (for
 $\rho_m,~\Lambda>0$)
\begin{eqnarray}
x=\frac{\kappa\sqrt{\rho_m}}{\sqrt{3}H},~~y=\frac{\sqrt{\Lambda}}{\sqrt{3}H}
\end{eqnarray}
From the above quantities, one obtains $x'$ and $y'$ as ($N\equiv \ln a$)
\begin{eqnarray}
x'&=&\frac{dx}{dN}=\frac{3}{2}(1+\omega_m)x(x^2-1)\\
y'&=&\frac{dy}{dN}=\frac{3}{2}(1+\omega_m)yx^2
\end{eqnarray}
The critical points, \emph{i.e}., the solutions corresponding to $x'=0,~y'=0$ and
the eigenvalues for the critical points \cite{copeland} are given by
\begin{eqnarray}
x_c&=&0,~~y_c= 1~(c.c.~dominant),~~~~~~~
\mu_{1}=0,~~\mu_{2}=-\frac{3}{2}(1+\omega_m)\label{ccd}\\
x_c&=& 1,~~y_c=0~(matter~dominant),~~~~~~~
\mu_{1}=3(1+\omega_m),~~\mu_{2}=\frac{3}{2}(1+\omega_m)
\end{eqnarray}
From these eigenvalues  we see that the
matter dominant phase is unstable for $\omega_{m}>-1$ while c.c. dominant phase has a
decaying mode, \emph{i.e}., $\mu_2<0$ being stable. (The existence
of a zero eigenvalue ($\mu_1=0$) in Equation~(\ref{ccd}) is originated from
the fact that two variables $x$ and $y$ are connected by the
relation $x^2+y^2=1$. Therefore in this case one can reduce to
one$-$dimensional space \cite{Alimohammadi}.) It
is well-known that there exists the non-perturbative solution of
Equations~(\ref{h1})--(\ref{h2}) which connects these two critical points as
\begin{eqnarray}
 a(t)=a_*(\sinh[At])^{2/3(1+\omega_m)},~~~H(t)=\sqrt{\frac{\Lambda}{3}}\coth[At],~~~
\rho_m=\frac{\Lambda}{\kappa^2}(\sinh[At])^{-2}\label{s2}
\end{eqnarray}
 where $A=(1+\omega_m)\sqrt{3\Lambda}/2$. This solution describes a smooth transition
 from matter-dominated power law expansion at early times into cosmological constant-dominated  exponential acceleration at late times.

 To check the stability of the above solutions Equation~(\ref{s2}),
 we consider the variation of \mbox{Equations~(\ref{h1})--(\ref{h2})}, which yields
\begin{eqnarray}
2H\delta{H}&=&\frac{\kappa^2}{3}\delta\rho_m\label{eomh}\\
\delta\dot{H}&=&-\frac{\kappa^2}{2}(1+\omega_m)\delta{\rho}_m\label{eomdoth}\\
\delta\dot\rho_m&+&3(1+\omega_m)\rho_m\delta
H+3(1+\omega_m)H\delta\rho_m=0\label{eomrho}
\end{eqnarray}
 Then we find the solution for the above Equations (\ref{eomh})--
 (\ref{eomrho}) as
\begin{eqnarray}
\delta H=
BC \tanh[At]e^{-f(t)}/\kappa^2,~~~ \delta\rho_m=C
e^{-f(t)}/\kappa^4
\end{eqnarray}
 where $B=(1+\omega_m)/4A$, $C$ is an arbitrary constant and $f(t)=-2At+\ln[-1+e^{4At}]+2\ln[\tanh[At]]$.\\
 Note that $f(t)$ approaches $2At$ as $t\rightarrow\infty$
 and both $\delta H$ and $\delta \rho_m$ decay, which implies that the
 solution~(\ref{s2}) is stable.

 Introducing the dimensionless density parameters $\Omega_m=\kappa^2\rho_m/3H^2$ and
$\Omega_{\Lambda}=\Lambda/3H^2$, Equation~(\ref{h1}) becomes
\begin{eqnarray}
\Omega_m+\Omega_{\Lambda}=1
\end{eqnarray}
Particularly for the solution
(\ref{s2}), $\Omega_m$ and $\Omega_{\Lambda}$ are given by
\begin{eqnarray}
\Omega_m=\frac{1}{(\cosh[At])^2},~~~
\Omega_{\Lambda}=1-\frac{1}{(\cosh[At])^2}
\end{eqnarray}
By using the experimental data, \emph{i.e}., $\rho_{\Lambda}=\frac{\Lambda
c^2}{8\pi G}\sim (10^{-12}Gev)^4\sim 10^{-8}erg/cm^3$, $H_0\sim
2.28\times 10^{-18}s^{-1}$ and the solution
$H_0=\sqrt{3\Lambda}\coth[At_0]/3$, we can evaluate the value
$At_0=1.27$. From these values we find
$\Omega_{\Lambda}$ and $t_0$ for the dust-like matter ($\omega_m=0$)
as
\begin{eqnarray}
\Omega_{\Lambda}\approx 0.73,~~~ t_0\approx 4.34\times 10^{17}s
\end{eqnarray}
which are in agreement with the observational data \cite{komatsu}.

\section{de Sitter Nonlinear Sigma Model with Potential}

In this section, we extend the analysis performed in the case of
$\Lambda-$CDM to the de Sitter nonlinear sigma model \cite{Lee} with a
potential term. The cosmological fluid with the equation of state
$\omega_f$ is also added. The starting action is
\begin{eqnarray}\label{des1}
S=\int \,d^4x
\sqrt{-g}~[~\frac{1}{2\kappa^2}R-\frac{g^{\mu\nu}}{\lambda^2}G_{\alpha\beta}(\Phi)
\partial_{\mu}\Phi^{\alpha}\partial_{\nu}\Phi^{\beta}-\epsilon V(\phi)+ {\cal L}_{fluid}~]
\end{eqnarray}
where $\Phi^{\alpha}=(\phi,\sigma^{i})~(i=1,2,3)$, $\lambda$ is a
dimensionless coupling constant and $\epsilon=\pm1$. Here
$G_{\alpha\beta}$ is the metric of the de Sitter target space,
\begin{eqnarray}
G_{\alpha\beta}=(-1,~+e^{2\kappa\xi\phi},~+e^{2\kappa\xi\phi},~+e^{2\kappa\xi\phi}~)
\label{de1}
\end{eqnarray}
where $\xi$ is an arbitrary positive constant and
 $V(\phi)$ is the potential given by
\begin{eqnarray}
V(\phi) = V_0 \exp(-\kappa\gamma\phi)
\end{eqnarray}
with an arbitrary constant $\gamma$ and $V_0 > 0$. Among diverse possibilities, we have chosen
exponential potential. A couple of  reasons could be cited. The first one is that this
potential is the prototype which gives rise to accelerating Universe and has interesting properties like scaling solution and attractor in the quintessence
 \cite{cope}
 or phantom model \cite{copeland}.
The second one is that it can yield non-perturbative solution of the
type discussed in the previous section in our case for the negative
potential  with $\epsilon=-1$. We regard the second and third
terms of Equation~(\ref{des1}) as representing the  dark energy sector.
For the matter part,  we assume a cosmological fluid of the
perfect fluid form, $T^{\mu}_{\nu}=(-\rho_f,~p_f,~p_f,~p_f)$, which
satisfies the continuity equation, $\nabla_{\mu} T^{\mu\nu}=0.$

We first note that the following ansatz
\begin{eqnarray}
\sigma^{i}=x^{i}\label{sol1}
\end{eqnarray}
solves the $\sigma^i$ field equations. The above ansatz first appeared in higher
 dimensional gravity theory in association with spontaneous
 compactification of the extra dimensions \cite{omero, gellmann,gellmann2,gellmann3}.
  It was revived in four dimensions recently in describing the accelerating Universe with the de Sitter nonlinear sigma model~\cite{Lee}.
Note that it does not break the isotropy and homogeneity of the universe
 as long as we do not introduce the potential for the $\sigma$
 fields.
 With this ansatz, the standard space-time metric Equation~(\ref{mmetric}),
 and $\phi=\phi(t)$, the evolution equations, are given by
\begin{eqnarray}
H^2&=&\frac{2\kappa^2}{3\lambda^2}~\Big[~-\frac{1}{2}\dot{\phi}^2+\frac{3}{2\kappa^4a^2}e^{2\kappa\xi\phi}+
\frac{\epsilon\lambda^2}{2} V(\phi)~\Big]+\frac{\kappa^2}{3}\rho_{f}\label{hh1}\\
\dot{H}&=&-\frac{\kappa^2}{\lambda^2}\Big[~-\dot{\phi}^2+\frac{1}{\kappa^4a^2}e^{2\kappa\xi\phi}~\Big]
-\frac{\kappa^2}{2}(1+\omega_{f})\rho_{f} \label{hh2}\\
0&=&\ddot{\phi}+3H\dot{\phi}-3\xi\frac{e^{2\kappa\xi\phi}}{\kappa^3a^2}+\frac{\epsilon\kappa\gamma\lambda^2}{2}
V_0 e^{-\kappa\gamma\phi}\label{hh3}
\end{eqnarray}
 and the continuity equation  implies $\rho_f
=\rho(0)[a(t)/a(0)]^{-3(1+\omega_f)},$ where $\omega_f$ is a
barotropic equation of state with $0<\omega_f<1$. We mention a
couple of properties of the above evolution equations.
The first one is that in spite of the $\phi$ being a phantom,
 the presence of the second term in Equation~(\ref{hh2}) that comes from the spatial
variations of $\sigma^i$ fields could prevent the Big Rip
singularity from  being developed;  it is not guaranteed that $\dot{H}$ will stay always positive
at late times when $\rho_f$ is ignored.
The other is that even in the $\epsilon=-1$ case, the weak energy
condition could not be violated (in the ordinary phantom model
with a negative potential, the weak energy condition is always
violated because of $\rho_{\phi}=-\dot{\phi}^2/2-V(\phi)<0$). This
again is due to the the second terms in Equations~(\ref{hh1}) and (\ref{hh2})
coming from the spatial variations.
In fact,  we will show that for $\epsilon=-1$ there exists an exact
cosmological solution of the type discussed in $\Lambda-$CDM case,
which interpolates between matter-dominated and dark energy-dominated epochs.

\section{Stability}

To perform the stability analysis and figure out the energy dominance of the
kinetic ($x$), spatial ($y$) and potential ($z$) parts in Equation~(\ref{hh1}), we first introduce the following dimensionless quantities,
\begin{eqnarray}
x\equiv\frac{\kappa\dot{\phi}}{\sqrt{3}\lambda H},
~~y\equiv\frac{e^{\kappa\xi\phi}}{\kappa \lambda
Ha},~~z\equiv\frac{\kappa\sqrt{V}}{\sqrt{3} H}\label{dimless}
\end{eqnarray}
Then the constraint Equation (\ref{hh1}) is given by
\begin{eqnarray}
 -x^2+ y^2+\epsilon z^2+\frac{\kappa^2\rho_f}{3H^2}=1
\end{eqnarray}
With $N=\ln{a}$, we obtain
\begin{eqnarray}
x'&\equiv
&\frac{dx}{dN}\!=\!\frac{x}{2}[-3(1\!-\!\omega_f)x^2-(1+3\omega_f)y^2\!+\!3(-1\!+\!\omega_f)
-3\epsilon(1\!+\!\omega_f)z^2]-\frac{3\epsilon\bar{\gamma}}{\sqrt{6}}z^2\!+\!
\bar{\xi}\sqrt{6} y^2\label{delx}\\
y'&\equiv
&\frac{dy}{dN}=\frac{y}{2}[-3(1-\omega_f)x^2-(1+3\omega_f)y^2\!+(1+3\omega_f)
-3\epsilon(1+\omega_f)z^2+2\bar{\xi}\sqrt{6} x]\label{dely}\\
z'&\equiv
&\frac{dy}{dN}=\frac{z}{2}[-3(1-\omega_f)x^2-(1+3\omega_f)y^2+3(1+\omega_f)
-3\epsilon(1+\omega_f)z^2-\sqrt{6}\bar{\gamma} x]\label{delz}
\end{eqnarray}
where
$\bar{\gamma}=\lambda\gamma/\sqrt{2},~\bar{\xi}=\lambda\xi/\sqrt{2}$.
The various   critical
points and their properties including the stabilities of the above equations are summarized in Table 1 with their eigenvalues  being given as follows:
\begin{itemize}
{\item point A}
\begin{eqnarray}
\hspace*{-20em}\mu_{1}=\frac{3}{2}(\omega_f-1),~~\mu_{2}=\frac{1+3\omega_f}{2},~~\mu_{3}=\frac{3(\omega_f+1)}{2}
\end{eqnarray}
{\item point B}
\begin{eqnarray}
&&\hspace*{-13.5em}\mu_{1,2}=-\frac{3(1-\omega_f)}{4}\pm
\frac{\sqrt{\bar{\gamma}^2(-1+\omega_f)(24+24{\omega_f}^2+7\bar{\gamma}^2+
\omega_f(48+9\bar{\gamma}^2))}}{4\bar{\gamma}^2}\\
&&\hspace*{-12.8em}\mu_3=\frac{\bar{\gamma}+3\omega_f\bar{\gamma}+6(1+\omega_f)\bar{\xi}}{2\bar{\gamma}}
\end{eqnarray}
{\item point C}
\begin{eqnarray}
&&\hspace*{-13.5em}\mu_{1,2}=-\frac{3(1-\omega_f)}{4}\pm
\frac{\sqrt{\bar{\xi}^2(-1+\omega_f)(4(1+3\omega_f)^2+3(5+27\omega_f)
\bar{\xi}^2)}}{4\bar{\xi}^2}\\
&&\hspace*{-12.8em}\mu_3=\frac{\bar{\gamma}+3\omega_f\bar{\gamma}+6(1+\omega_f)\bar{\xi}}
{4\bar{\xi}}
\end{eqnarray}
{\item point D}
\begin{eqnarray}
\hspace*{-16em}\mu_{1}=-\frac{6+\bar{\gamma}^2}{2},
~~\mu_{2}=-3-3\omega_f-\bar{\gamma}^2,
~~\mu_{3}=-\frac{2+\bar\gamma^2+2\bar\gamma\bar\xi}{2}
\end{eqnarray}
{\item point E}
\begin{eqnarray}
\hspace*{-16em}\mu_{1}=-1-3\omega_f-6\bar{\xi}^2,~~
\mu_{2}=1-3\bar{\xi}^2-\frac{3\bar{\gamma}\bar{\xi}}{2},~~
\mu_{3}=-2-3\bar{\xi}^2
\end{eqnarray}
{\item point F}
\begin{eqnarray}
&&\hspace*{-14.5em}\mu_{1,2}=-\frac{\bar{\gamma}+3\bar{\xi}}{\bar{\gamma}+2\bar{\xi}}
\pm\frac{\sqrt{-(8-6\bar{\gamma}\bar{\xi}(\bar{\gamma}
+2\bar{\xi})^2)-3\bar{\gamma}^2+10\bar{\gamma}\bar{\xi}+33\bar{\xi}^2}}
{\bar{\gamma}+2\bar{\xi}}\\
&&\hspace*{-13.5em}\mu_{3}=-\frac{\bar{\gamma}+3\omega_f\bar{\gamma}+6(1+\omega_f)\bar{\xi}}{\bar{\gamma}+2\bar{\xi}}
\end{eqnarray}
\end{itemize}
Note that all of the above eigenvalues of the critical points do not have any $\epsilon$-dependence even though the
various critical points themselves carry its dependence.

\begin{table}[H]
\centering
\caption[crit]{The classification and the properties of the critical points.}
\label{}
\includegraphics[width=0.95\textwidth]{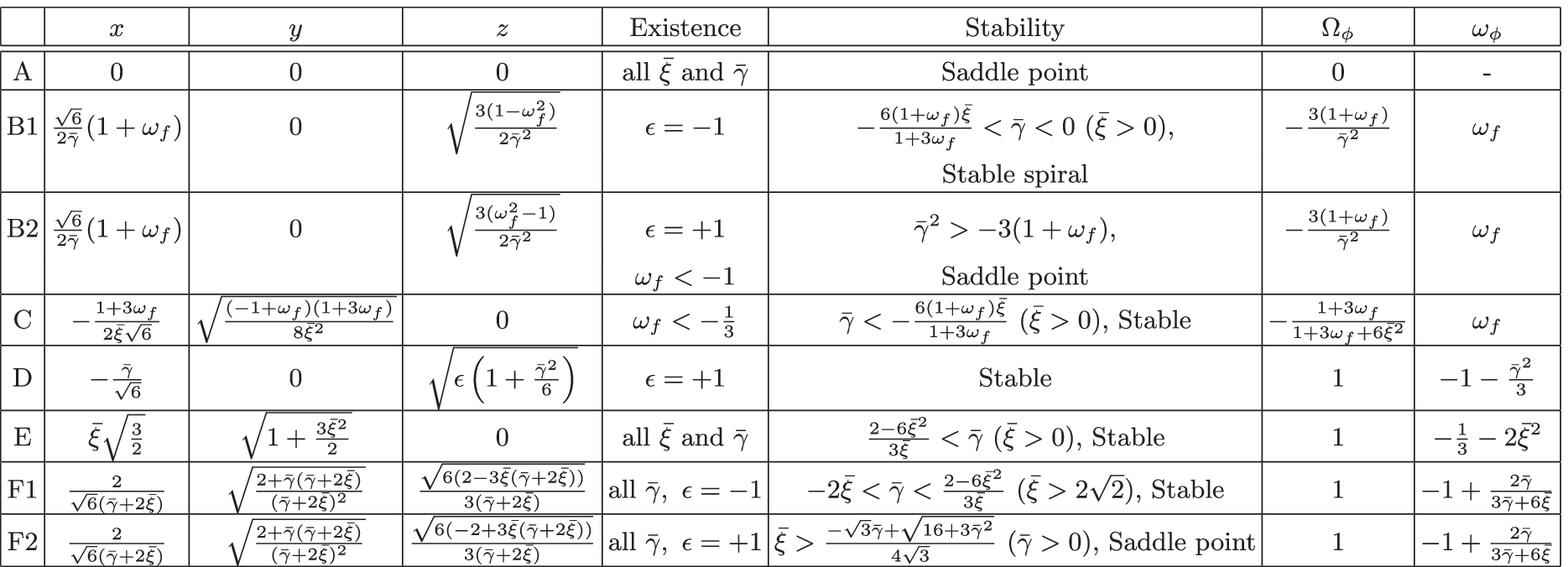}
\end{table}

In Table 1, the second to fourth columns denote energy contents of the dark energy sector, the fifth column shows
the conditions for the existence of the solutions of the critical Equations (\ref{delx})--(\ref{delz}).
The critical points B and F are further divided for $\epsilon=\pm 1$.
The sixth column is about the stability condition for each eigenvalue. In the seventh column, $\Omega_\phi=-x^2+y^2+\epsilon z^2$
is the density parameter for the dark energy, and the last column displays the equation of state for the dark energy.
First, we note that the results of phases A, B2, D with $\epsilon=+1$ and $y=0$ agree
with those in the literature~\cite{copeland}, which correspond to the phantom with a positive potential.
In particular, the system approaches the scalar-dominated solution D with non-relativistic dark matter $(\omega_f=0)$.
The point F2 also corresponds to scalar-dominated solution but with a non-zero value of $y$.

Let us focus on the $\epsilon=-1$
case from here on. Then, points E and F1 both give scalar-dominated solutions. Which solution will the system
choose depends on the allowed parameters for the existence and stability conditions and on the initial conditions.
For solution E, $z=0$ and $\omega_{\phi}$ approaches to $-1$ when $\bar\xi^2\rightarrow 1/3$.
On the other hand, for solution F, all $x$, $y$ and $z$ contribute to the dark energy and $\omega_{\phi}$ converges to $-1$ when $\bar{\xi}$ goes to
infinity.
In the next section, we present an explicit solution of Equations~(\ref{hh1})--(\ref{hh3}),
which prolongs from B1 at early times to E at late times.



\section{Exact Cosmological Solution}

In order to investigate the exact solution for
Equations~(\ref{hh1})--(\ref{hh3}), we first assume that
$a=e^{\kappa\xi\phi}/\sqrt{f}$, which corresponds to
$H=\kappa\xi\dot\phi$. Here $f$ is an arbitrary constant at this stage.
In the case of $\epsilon=-1$, from the above
ansatz one can find an exact solution as follows:
\begin{eqnarray}
\phi(t)&=&
\phi(0)+\frac{\sqrt{f}}{\kappa^2}t+\frac{2}{3\kappa\xi(1+\omega_f)}
\ln\left[\frac{1+C
e^{-3\xi(1+\omega_f)t/\kappa}}{1+C}\right]\label{solphi}
\end{eqnarray}
where $C$ is an arbitrary constant and $V_0$ and $\rho_0$ are given by
\begin{eqnarray}\label{vsol5}
V_0&=&\frac{f(1-\omega_f)}{\kappa^4\lambda^2(1+\omega_f)}
e^{3(1+\omega_f)
\kappa\xi\phi(0)}\left\{\frac{(1+C)^2}{-4C}\right\}\\
\rho_0&=&\frac{-4Cf}{\kappa^4(1+C)^2}\left\{3\xi^2+2\frac{1}{(1+\omega_f)\lambda^2}\right\}
\label{rhosol5}
\end{eqnarray}
with $\xi^2=2/3\lambda^2$, $\gamma=3(1+\omega_f)\xi$. It is to be noticed
that $C$ should be a negative value to preserve the weak energy
condition. In this case, introducing a time scale defined by $\vert
C\vert =e^{-3\frac{\xi}{\kappa}(1+\omega_f) \sqrt{f}t_{*}}$, Equation~(\ref{solphi}) and $a=e^{\kappa\xi\phi}/\sqrt{f}$ reduce to
\begin{eqnarray}\label{phisol5}
\phi(t)&=& \frac{2}{3\kappa\xi(1+\omega_f)}
\ln\left(\sinh\left[\frac{3(1+\omega_f)}{2}
\frac{\xi}{\kappa}\sqrt{f}(t+t_*)\right]\right)+\tilde{\phi}(0)\\
a(t)&=&\tilde{a}\sinh\left[\frac{3(1+\omega_f)}{2}
\frac{\xi}{\kappa}\sqrt{f}(t+t_*)\right]^{\frac{2}{3(1+\omega_f)}}
\label{asol5}
\end{eqnarray}
where
$\tilde{\phi}(0)=\phi(0)-2\ln\Big[\sinh(\frac{3(1+\omega_f)}{2}\frac{\xi}{\kappa}\sqrt{f}
t_{*})\Big]/(3(1+\omega_f)\kappa\xi)$ and
$\tilde{a}=e^{\kappa\xi\tilde{\phi}(0)}/\sqrt{f}$

We note that for $t\to -t^{*}$ the solution (\ref{asol5}) behaves as
$a\sim t^{2/(3(1+\omega_f))}$, which shows that it represents an
expanding Universe with cosmological fluid $\omega_f$, whereas at
late times, it represents  an accelerating Universe with $a(t)\sim
e^{\xi\sqrt{f}t/{\kappa}}$. Now, let us check that  the above
solution indeed corresponds to the one that prolongs from B1 at
early times to E  at late times, as was mentioned in the
previous section. To this end we first substitute the solutions
(\ref{vsol5})--(\ref{asol5}) into the dimensionless quantities
$x,~y,~z$ of  Equation~(\ref{dimless}). Then one can find
\begin{eqnarray}\label{xfinal}
x&=&\frac{1}{\sqrt{3}\lambda\xi}\\
y&=&\frac{1}{\lambda\xi}\frac{\sinh[\frac{3(1+\omega_f)}{2}\frac{\xi}{\kappa}\sqrt{f}(t+t_{*})]}
{\cosh[\frac{3(1+\omega_f)}{2}\frac{\xi}{\kappa}\sqrt{f}(t+t_{*})]}
\label{yfinal}\\
z&=&\frac{1}{\sqrt{3}\lambda\xi}
\frac{(\frac{1-\omega_f}{1+\omega_f})^{\frac{1}{2}}}{\cosh[\frac{3(1+\omega_f)}{2}\frac{\xi}{\kappa}\sqrt{f}(t+t_{*})]}
\label{zfinal}
\end{eqnarray}
When choosing $\xi=\sqrt{2}/\sqrt{3}\lambda (\bar{\xi}=1/\sqrt{3})$,
 we see that at early times, \emph{i.e}.,
$t\to -t^{*}$, the quantities $x,~y,~z$ of Equations~(\ref{xfinal})--(\ref{zfinal}) are given by
\begin{eqnarray}
x=\frac{1}{\sqrt{2}},~~y=0,~~z=\sqrt{\frac{1-\omega_f}{2(1+\omega_f)}}\nonumber
\end{eqnarray}
and at late times ($t\to \infty$) they become
\begin{eqnarray}
x=\frac{1}{\sqrt{2}},~~y=\sqrt{\frac{3}{2}},~~z=0\nonumber
\end{eqnarray}
If we identify $\bar\gamma=(1+\omega_f)\sqrt{3}$, and substituting $\bar{\xi}=1/\sqrt{3}$ into B1 and E of Table 1,
we exactly find the above values of $x$, $y$ and $z$.
Therefore, we conclude that  our solution corresponds to
the phase B1 at early times and  to the phase E at late times.

We now determine the allowed range of $\omega_f$ for this exact solution
by comparing with the current observational data.
It turns out that the non-relativistic dark matter with $\omega_f=0$
is not consistent with the current value of $t_0$.
To carry out
this in detail,  we recall the density parameter of the scalar
field given by (we set $t_{*}\sim 0$)
\begin{eqnarray}
\Omega_{\phi}&=&-x^2+y^2-z^2\nonumber\\
&=&(\tanh[Bt])^2-\frac{1}{(1+\omega_f)
(\cosh[Bt])^2}\label{omegaphi}
\end{eqnarray}
where $B=3(1+\omega_f)\xi\sqrt{f}/2\kappa$ with
$\xi=\sqrt{2}/\sqrt{3}\lambda$ and the Hubble parameter
\begin{eqnarray}
H&=&\frac{2B}{3(1+\omega_f)}\coth[B t]\label{Hphi}
\end{eqnarray}

We first comment on the dependence
of $\Omega_\phi$, and $H$ of Equations~(\ref{omegaphi}) and (\ref{Hphi}) on the initial conditions.
The exact solutions, Equations~(\ref{phisol5}) and (\ref{asol5}), show the dependence of $\phi$ and $a$ on the initial conditions.
However, the dependence of the quantities $x,~ y,~$ and $z$ on the initial conditions are absent as can be seen in the expressions of Equations~ (\ref{xfinal})--(\ref{zfinal}).  This may be due to the fact that our exact cosmological solution
satisfies the condition $a=e^{e^{\kappa \xi\phi}}/\sqrt{f}$ and is of a non-perturbative nature. The only remaining dependence
is through the variable $t_*$, which we have set to zero without loss of generality. Therefore,  $\Omega_\phi$ and $H$ of
Equations~(\ref{omegaphi}) and (\ref{Hphi})
are insensitive to the initial conditions as far as the non-perturbative solutions (\ref{phisol5}) and (\ref{asol5}) are concerned.
This aspect is rather unexpected because there is  a priori no reason why the quantities $x$, $y$ and $z$ defined in Equation~(\ref{dimless})
should not depend on the initial conditions of $\phi$ and $a$.

In the above two Equations (\ref{omegaphi}) and (\ref{Hphi}), we treat  $\Omega_{\phi,0}$, $H_{0}$ at $t=t_0$
as input parameters to determine $\xi\sqrt{f}$ and especially $\omega_f$.
Since the observational data \cite{komatsu} $\Omega_{\phi,0}\simeq0.726\pm0.015,~H_{0}\simeq2.28\pm0.04\times 10^{-18}s^{-1}$ and $t_0\simeq4.33\pm0.04\times 10^{17}s$
are given with experimental uncertainty, these equations give permitted range of $\omega_f$.
We use the strategy that we first pick up a specific value of $\omega_f$. Then, Equations~(\ref{omegaphi}) and (\ref{Hphi})
will give allowed ranges of $B$ for $\Omega_{\phi,0}$ and $H_{0}$ respectively. We can plot
these quantities in ($B,t$)-plane. If there exist  intersecting region, then, this value of $\omega_f$ is allowed.
The results are displayed in  Figure 1. We find that for values of
$\omega_f< 0.13$ and  $\omega_f>0.22$, there does not exist a region
intersected by the bands of  $\Omega_{\phi,0},~H_{0}$ and $t_0$.
Therefore, we conclude that the allowed value $\omega_f$ consistent
with the current observational data is given by
 $0.13<\omega_f<0.22$,  which corresponds to an exotic cosmological fluid with non-vanishing pressure.


\begin{figure}[H]
\centering
\includegraphics[width=0.46\textwidth]{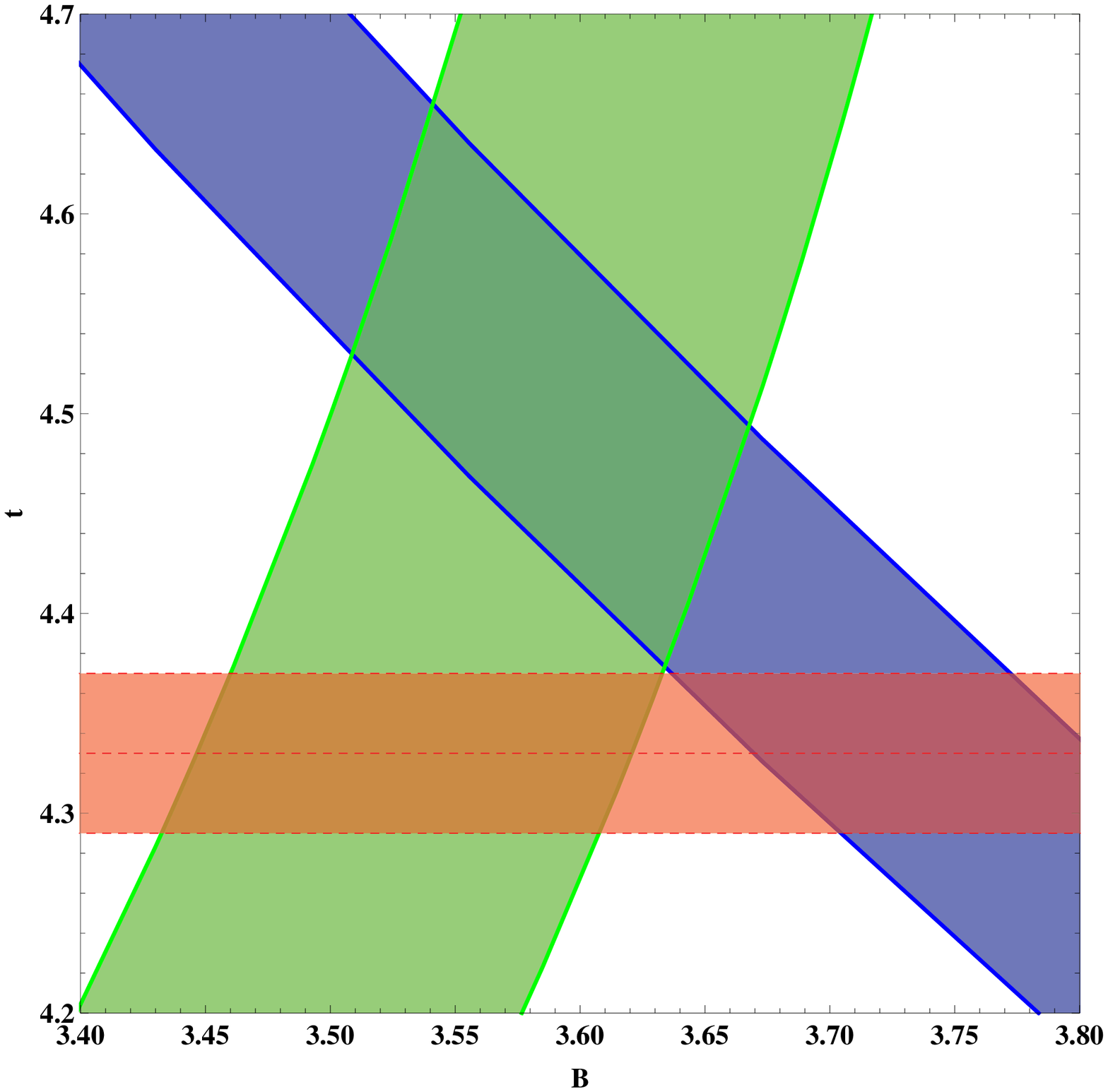}
\includegraphics[width=0.46\textwidth]{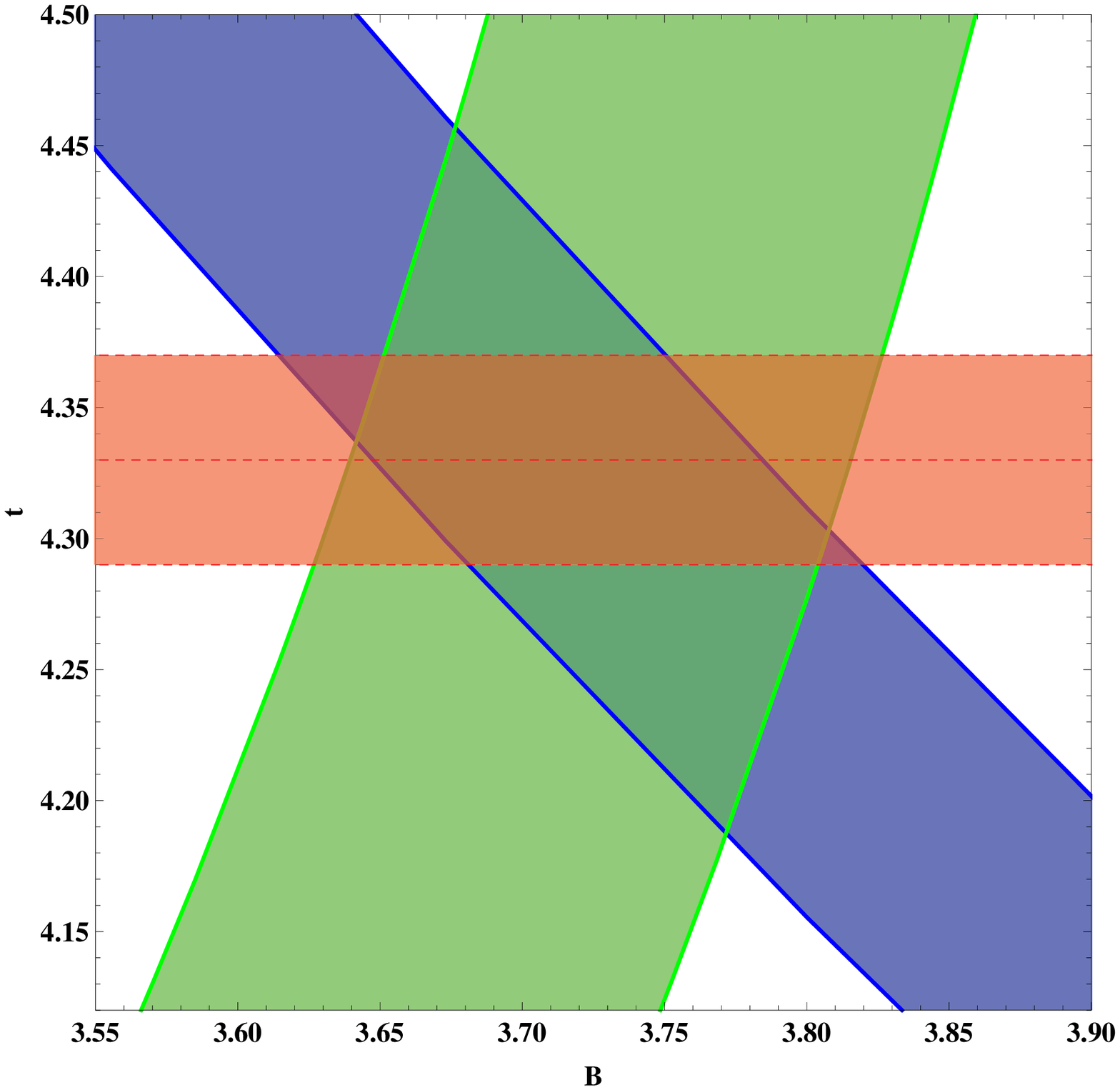}
\includegraphics[width=0.46\textwidth]{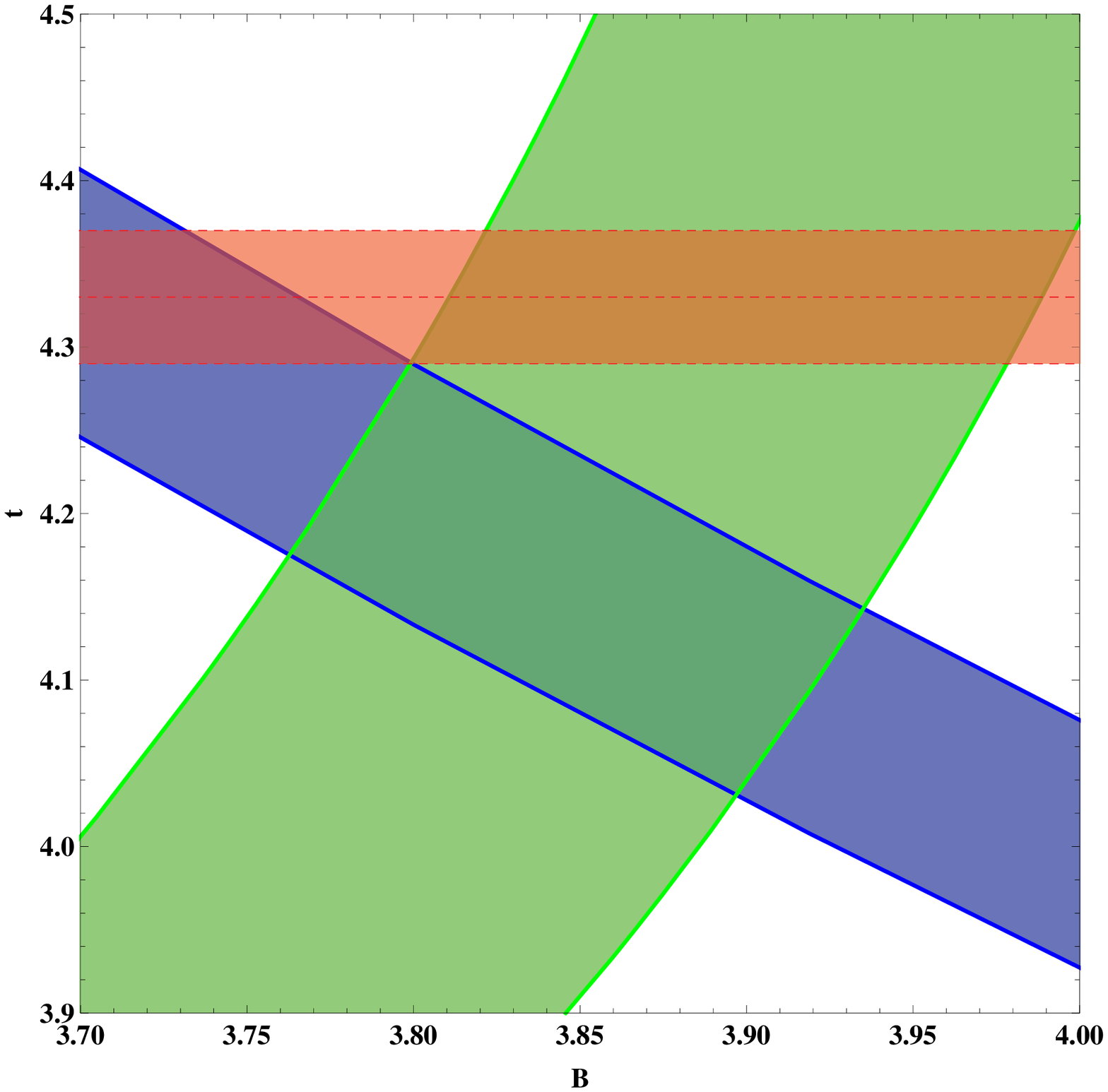}
\caption{The plot of $\Omega_{\Lambda,0}(blue)\simeq0.726\pm0.015,~H_{0}(green)\simeq2.28\pm0.04\times 10^{-18}s^{-1}$
on the ($B$, $t$)-plane for $\omega_f=0.13$(left),~ $\omega_f=0.18$(right),
~$\omega_f=0.22$(lower). The red band is the current age with uncertainty $t_0\simeq4.33\pm0.04\times 10^{17}s$.}
\label{fig:mt2jetdif}
\end{figure}

\section{Conclusion}

In this paper, we have investigated various cosmological
consequences of the de Sitter nonlinear sigma model with the
exponential potential and cosmological fluid term. It consists of a
phantom and triplet scalar fields. We have displayed that the
presence of the triplet $\sigma$ fields with isotropic spatial
dependence provides  contrasting features to the phantom model
without such fields. For example, the energy and pressure coming
from the spatial variations could prevent  the Big Rip singularity
from being developed. In particular, we found an exact cosmological
solution in the case of a negative potential. In this solution, the
Universe undergoes a power law expansion at early times as in
$\Lambda-$CDM. But the difference is that unlike the $\Lambda-$CDM we
have a non-vanishing scalar energy contribution ($x\neq 0, z\neq 0$
of B1 phase) even in the matter dominated epoch. At late times, a
complete dark energy-dominance is achieved with $\Omega_\phi=1$ and
$\omega_\phi=-1$ with a suitable choice of the parameter.

The Hubble parameter Equation~(\ref{Hphi}) is exactly the same as $H$ of
$\Lambda-$CDM in Equation~(\ref{s2}) when we identify the parameter  as
$\xi\sqrt{3f}/\kappa=\sqrt{\Lambda}$. However, with this choice, one
can check that the energy density $\rho_f$ of cosmological
fluid is different from the matter density $\rho_m$ of Equation~(\ref{s2})
in $\Lambda-$CDM, and is given by
$\rho_{f}/\rho_{m}=1+1/(\omega_{f}+1)$. Note that for   $0
\leq\omega_f\leq 1$, $\rho_f$ is always greater than $\rho_m$. Also,
the dark energy density from Equation~(\ref{omegaphi}) approaches
$\Lambda/\kappa^2$ asymptotically from below.
 These show that even if the scale factors in both cases behave exactly the same,
there exist qualitative differences between the two approaches.

We also have  shown that to be consistent with the observational
data, the equation of state for the cosmological fluid has to be
within the range $0.13<\omega_f<0.22$ unlike $\Lambda-$CDM with a
dust like matter. This range of $\omega_f$ appears in the literature
\cite{exotic,exotic2} corresponding to cosmic strings with
$-1/3<\omega_f<1/3$  or domain walls with $-2/3<\omega_f<1/3$.
We notice that in $\Lambda-$CDM model, the dark
matter has the perfect fluid form and admits a barotropic equation of
state. In this case, the dark matter should be pressureless in order
 to be in accordance with
the observational data.
However, it is known \cite{Bharadwaj,Su} that the dark matter can also be
described by an anisotropic fluid but only in the case of non-zero effective pressure
 or a polytropic equation of state
\cite{Boehmer}. On
the other hand, our exotic cosmological fluid  not only satisfies
a barotropic equation of state but also takes a perfect fluid form.
In spite of this, it should have non-zero pressure to be in
agreement with the observational data.

We conclude with the following remark.  We found that with $\omega_f=0$ the evolution is exactly that of $\Lambda$-CDM. But this does not fit the observational data as was discussed, e.g., $\omega_f$ has to be in the range of 0.13 and 0.22 to give the observed age of the universe. We speculate that this feature of deviation from the $\Lambda$-CDM for general time-varying dark energy
density holds in general. That is, in models where dark energy density varies, the dark matter with $\omega_m=0$
might have some difficulty in fitting the observations. Nevertheless, this does not imply that the dynamical dark energy models
with non-zero equation of state dark matter must be excluded,
and it remains to be seen whether the exotic cosmological fluid considered in this work could be related to the realistic dark matter  candidates.


\section*{Acknowledgments}
\vspace{12pt}

We like thank Joohan Lee and Tae Hoon Lee for useful discussions.
This work was supported by the Basic Science Research Program through the National Research Foundation of Korea (NRF) funded by the MEST (2011-0026655) and
by NRF grant funded by the Korea government (MEST) through the Center for Quantum Spacetime (CQUeST) of Sogang University with grant number 2005-0049409.

\bibliographystyle{mdpi}
\makeatletter
\renewcommand\@biblabel[1]{#1. }
\makeatother

\end{document}